\documentstyle[12pt]{article}

\begin{document}

\baselineskip=22pt plus 1pt minus 1pt

\begin{center}{\large \bf
$\Delta I=4$ and $\Delta I=8$ bifurcations
in rotational bands of diatomic molecules}

\bigskip\bigskip

{Dennis Bonatsos$^{*\#}$,
C.~Daskaloyannis$^+$,
S. B. Drenska$^\dagger$,
G. A. Lalazissis$^+$,
N. Minkov$^\dagger$,
P. P. Raychev$^\dagger$,
R. P. Roussev$^\dagger$
\bigskip

{$^*$ European Centre for Theoretical Studies in Nuclear Physics
and Related Areas (ECT$^*$)}

{Strada delle Tabarelle 286, I-38050 Villazzano (Trento), Italy}

{$^{\#}$ Institute of Nuclear Physics, N.C.S.R.
``Demokritos''}

{GR-15310 Aghia Paraskevi, Attiki, Greece}

{$^+$ Department of Physics, Aristotle University of
Thessaloniki}

{GR-54006 Thessaloniki, Greece }

{$^\dagger$ Institute for Nuclear Research and Nuclear Energy, Bulgarian
Academy of Sciences }

{72 Tzarigrad Road, BG-1784 Sofia, Bulgaria}}
\end{center}

\bigskip\bigskip
\centerline{\bf Abstract} \medskip
It is  shown that the recently observed
$\Delta I=4$ bifurcation seen in superdeformed nuclear bands
is also occurring in
 rotational bands of diatomic
molecules. In addition, signs of a $\Delta I=8$ bifurcation, of the same order
of magnitude as the $\Delta I=4$ one, are observed both in superdeformed
nuclear bands and rotational bands of diatomic molecules.

\bigskip\bigskip
PACS numbers: 33.10.Ev, 21.10.Re, 21.60.Ev

\newpage

Rotational bands of diatomic molecules \cite{Herzberg} and rotational bands
of deformed nuclei \cite{BM} have many features in common, despite the
different energy scales involved in each case. Molecular rotational bands
are in general closer to the behavior of the rigid rotator than their nuclear
counterparts. In the last decade much
interest has been attracted by superdeformed nuclear bands \cite{Twin,R1,R2},
which are characterized by relatively high angular momenta and behavior
closer to the rigid rotator limit in comparison to normal deformed nuclear
bands.

A rather surprising feature has been recently discovered
\cite{Fli,Ced} in superdeformed
nuclear bands: Sequences of states differing by four units of angular
momentum are displaced relative to each other, the relative shift being
of order of $10^{-4}$ of the energies separating the levels of these bands.
A few theoretical proposals for the possible explanation of this bifurcation,
$\Delta I =4$ which is also called $\Delta I=2$ staggering, have
already  been made \cite{HMott,Macc,PavFli,MQ}

A reasonable question is therefore whether $\Delta I=4$
bifurcations (i.e. $\Delta I=2$ staggering) also occur
in rotational spectra of diatomic molecules.
We are going to show in this work
that this is indeed the case. Bifurcations with $\Delta I=8$ and
$\Delta I=12$ will also be discussed.

In nuclear physics the experimentally determined quantities are the
$\gamma$-ray transition energies between levels differing by two units
of angular momentum ($\Delta I=2$). For these the symbol

\begin{equation}
 E_{2,\gamma}(I) = E(I+2)-E(I)
\label{eq:1old}
\end{equation}
is used, where $E(I)$ denotes the energy of the level with angular momentum
$I$.
The deviation of the $\gamma$-ray transition energies from the
rigid rotator behavior can be measured by the quantity
\cite{Ced}

\begin{equation}
 \Delta E_{2,\gamma}(I) = {1\over 16} (6E_{2,\gamma}(I) -4E_{2,\gamma} (I-2)
-4E_{2,\gamma}(I+2) +E_{2,\gamma}(I-4) +E_{2,\gamma}(I+4)).
\end{equation}

\noindent Using the rigid rotator expression $E(I)=A I(I+1)$ one
can easily see that
in this case $\Delta E_{2,\gamma} (I) $ vanishes.
This is due to the fact that Eq. (2) is the
discrete approximation of the fourth derivative of the function
$E_{2,\gamma}(I)$.

Several nuclear superdeformed rotational bands such as (a) to
(e) for  $^{149}$Gd
\cite{Fli} and the bands (1) to (3) for $^{194}$Hg \cite{Ced}
were analyzed. The corresponding tables are not included in
this short presentation,
being reserved for a forthcoming longer publication.
The analysis shows that the $\Delta E_{2,\gamma} (I)$ values
exhibit an anomalous staggering. It should be noted, however,
that only for the band (a)
of $^{149}$Gd \cite{Fli}, the amplitude of the oscillations
(see for example fig. 3 of ref \cite{Ced}) is definitely
outside the experimental errorbars. The following observations can be made:

\noindent i) $\Delta E_{2,\gamma} (I)$ obtains alternating
positive and  negative
values. This is why this effect has also been called ``$\Delta I=2$
staggering''.

\noindent ii) The magnitude of $\Delta E_{2,\gamma}(I)$ is of
order $10^{-4}$--$10^{-5}$
of that of the $\gamma$-ray transition energies.

\noindent iii) The staggering oscillation width is an increasing
function  of the
angular momentum $I$.

In the case of molecules the experimentally determined
quantities regard the R branch ($I\rightarrow I+1$) and the P branch
($I\rightarrow I-1$). They are related to transition energies through the
equations \cite{Barrow}
\begin{equation}
E^R(I)-E^P(I)= E_{v=1} (I+1) -E_{v=1} (I-1) = DE_{2, v=1} (I-1),
\end{equation}
\begin{equation}
E^R(I)-E^P(I+2) = E_{v=0}(I+2)-E_{v=0}(I)=DE_{2, v=0}(I),
\end{equation}
where in general
\begin{equation}
 DE_2 (I) = E(I+2)-E(I),
\end{equation}
and $v$ is the vibrational quantum number. $\Delta I=2$ staggering  can then
be estimated by using Eq. (2), with $E_{2,\gamma}(I)$ replaced by $DE_2(I)$:
\begin{equation}
 \Delta E_2 (I)= {1\over 16} (6 DE_2(I)-4 DE_2(I-2)-4 DE_2(I+2) +DE_2(I-4)
+DE_2(I+4)).
\end{equation}
It is noted, that for the sake of simplicity a normalized form
of the discrete fourth derivative is used in (6) as well as in
the subsequent equations : (7), (9), and (12).
We have analyzed quite a few molecular rotational bands
for several diatomic molecules. Some of them,   revealing a
staggering effect, are shown in Figs 1, 2 and 3.
In Fig. 1  a typical example of the $\Delta I=2$
staggering in molecular rotational spectra is shown. The (1-1), $v$=1
rotational band of the   C$^1\Sigma^+$--X$^1\Sigma^+$ system of YD
(data from \cite{RamYD})  has been used.
In Fig. 2 the (2-2), $v=$1
rotational band of the
A$^1\Sigma^+$--X$^1\Sigma^+$ system of YN (data from \cite{RamYN}) is drawn.
In Fig. 3 the (4-3), $v$=1 band for the CS is given (data from \cite{RamCS}).
Similar results can be obtained from the other available bands of these
molecules and from the
A$^6\Sigma^+$--X$^6\Sigma^+$ system of CrD (data from \cite{RamCrD}).
The following comments are in place:

\noindent i) $\Delta E_2(I)$ exhibits alternating signs with
increasing $I$,  a
fingerprint of $\Delta I=2$ staggering for the $v=1$ band, while
the $v=0$ band data (which are not shown)
do not permit a clear cut identification of the staggering
effect. Even for the $v=1$ bands, only in the case of YD is the magnitude
of the effect clearly smaller than the experimental uncertainties.

\noindent ii) The magnitude of the perturbation, $\Delta E_2(I)$, is of order
$10^{-3}$--$10^{-5}$ of that of the interlevel separation energy.

\noindent iii) The staggering oscillation width is not a
monotonically  increasing
function of the angular momentum $I$.
The irregularities in the magnitude of $\Delta E_2(I)$ might indicate the
presence of subsequent bandcrossings \cite{MRM}.
It is known that the bandcrossing effect is seeing only when the interaction
between the two bands which cross each other is relatively weak
\cite{VDS}. Therefore only the levels neighboring the crossing are affected
by the interaction. From Eq. (6) it is then clear that perturbing an
energy level results in perturbing 5 consequent values of
$\Delta E_2(I)$. In view of this, Fig. 1 looks very much like depicting
two subsequent bandcrossings. Figures 2 and 3, however, do not immediately
accept such an interpretation. Bandcrossing has been recently suggested as
a possible source of the $\Delta I=4$ bifurcation in nuclei \cite{Sun,Reviol}.
Certainly more work is needed in this direction.

\noindent iv) The staggering effect is more prominent in the
case of  even angular
momentum data than in the case of odd angular momentum data
(which are not shown).

One might further wonder if bifurcations with $\Delta I >4$ can also occur.
In the nuclear case, the existence of $\Delta I =4$ staggering can be checked
by using the quantity

\begin{equation}
 \Delta E_{4,\gamma} (I)= {1\over 16} (6 E_{4,\gamma}(I) -4 E_{4,\gamma}(I-4)
-4 E_{4,\gamma}(I+4) +E_{4,\gamma}(I-8) +E_{4,\gamma}(I+8)),
\end{equation}
where
\begin{equation}
 E_{4,\gamma}(I)= E(I+4)-E(I).
\end{equation}
\noindent Results for several superdeformed nuclear bands have
been  calculated.
In Fig. 4 the $\Delta I=8$ bifurcation for the superdeformed
band (a) of $^{149}$Gd \cite{Fli} is shown. Note that no angular momentum
assignments are shown, since they are still uncertain.
The following remarks apply:

\noindent i) $\Delta E_{4,\gamma} (I)$ acquires alternating
signs with  increasing $I$,
indicating the existence of a $\Delta I=8$ bifurcation.

\noindent ii) The order of magnitude of the $\Delta I=4$
staggering is the same as
that of the $\Delta I=2$ staggering.

In the case of diatomic molecules one can search for $\Delta I =4$
staggering by using the quantity

\begin{equation}
 \Delta E_4 (I) = {1\over 16} (6 DE_4(I) -4 DE_4(I-4) -4 DE_4(I+4)
+DE_4(I-8) +DE_4(I+8)),
\end{equation}
where

\begin{equation}
 DE_4 (I)= E(I+4)-E(I).
\end{equation}

\noindent In our study we have analyzed the larger known bands of the
molecule CS, i.e. the bands (1-0) $v=1$, (2-1) $v=0$, (4-3) $v=0$,
(2-1) $v=0$ \cite{RamCS}.
The results
for the rotational bands of the
A$^6\Sigma^+$--X$^6\Sigma^+$ system of CrD (data from \cite{RamCrD}),
the C$^1\Sigma^+$--X$^1\Sigma^+$ system of YD
(data from \cite{RamYD}),
and the A$^1\Sigma^+$--X$^1\Sigma^+$ system of YN
(data from \cite{RamYN}) were also
considered. For these molecules  experimental data of long enough
bands exist, permitting the calculations.
In Fig. 5 the $\Delta I=4$ staggering ($\Delta I =8$
bifurcation)  for a $v=1$ band of the molecule YD
is shown. The following comments can be made:

\noindent i) Alternating signs of $\Delta E_4(I)$,
a fingerprint of $\Delta I=8$ bifurcation, are observed.

\noindent ii) The magnitude of the $\Delta I =4$ staggering appears to be
the same as that of the $\Delta I=2$ staggering.

$\Delta I=12$ bifurcation i.e. $\Delta I=6$ staggering can be
searched for  through use of the quantity
\begin{equation}
 \Delta E_6 (I)={1\over 16} (6DE_6(I)-4 DE_6(I-6)-4DE_6(I+6)+DE_6(I-12)
+DE_6(I+12)),
\end{equation}
where
\begin{equation}
 DE_6(I)=E(I+6)-E(I).
\end{equation}
 Calculations have been carried out for a few cases of  rotational
 bands of CS (data from \cite{RamCS}),
 and for the B$^1\Sigma_u^+$--X$^1\Sigma_g^+$
 system of $^{63}$Cu$^{65}$Cu (data from \cite{RamCu65}),
 in which bands long enough for such
a calculation are known. These results look like being in favor of the
existence of $\Delta I=6$ staggering of the
same order of magnitude as $\Delta I=4$ and  $\Delta I=2$
staggering, but they are not enough for drawing any final conclusions.

The observation of $\Delta I=2$ and $\Delta I=4$ staggering in rotational
spectra of diatomic molecules offers a corroboration of the existence of the
same effect in nuclei, since the experimental techniques used in each case
are quite different, so that the occurrence of the same systematic errors
in both cases is improbable. Furthermore, the energy scales involved in
nuclei and molecules are very different (the separation of energy levels
in molecules is of the order of 10$^{-2}$eV, while in nuclei of the order
of 10$^5$eV), but the staggering effects are
in both cases of the same order of magnitude relative to
the separation of the
energy levels, indicating that the same basic
mechanism, possibly related to some perturbations of given symmetry,
might be responsible for these effects in both cases.

\noindent In conclusion, we have shown that:

\noindent i) $\Delta I=2$ staggering, first observed in superdeformed
nuclear bands \cite{Fli,Ced},
occurs as well in rotational bands of diatomic molecules.

\noindent ii) In all cases the magnitude of the $\Delta I=2$ staggering is
$10^{-3}$--$10^{-5}$ of that of the separation of the energy levels.

\noindent iii) Furthermore $\Delta I=4$ staggering appears to be present both
 in superdeformed
nuclear bands as well as in rotational bands of diatomic molecules, its
order of magnitude being the same as that of the $\Delta I=2$ staggering
in the same physical system.

\noindent iv)
In most cases the magnitude of the staggering does not show a simple
dependence on angular momentum. In several cases one sees about
5 points deviating
very much for the smooth rotational behavior, then several points much
closer to the pure rotational behavior, then again about 5
points deviating  very
much from the smooth rotational behavior, and so on. Such a picture raises
suspicions for the presence of bandcrossings at the points at which the
large deviations occur \cite{Sun,Reviol}.

Concerning the theoretical explanation of the $\Delta I=2$ staggering effect,
some proposals in the nuclear physics framework already exist
\cite{HMott,Macc,PavFli,MQ}. However, one cannot draw any firm
conclusions up to now.
In view of the present results, further efforts,
in investigating other molecular and nuclear data, are necessary.

\bigskip\medskip

Two authors (DB and GAL) have been supported by the E.U. under contracts
ERBCHBGCT930467 and ERBFMBICT950216 respectively. Another author
(PPR) acknowledges support from the Bulgarian Ministry of Science
and Education under contracts $\Phi$-415 and $\Phi$-547.
Three authors (DB, CD, GAL) have been supported by the Greek Secretariat of
Research and Technology under contract PENED95/1981.

\vfill\eject

\newpage

\centerline{\bf  Figure Captions}
\bigskip\bigskip\bigskip

\begin{itemize}

\item[{\bf Fig. 1}] $\Delta I$ =2 staggering for the
rotational band (1-1) , $v$=1 of the
  C$^1\Sigma^+$--X$^1\Sigma^+$ system of the molecule YD.
Even values of $I$ are used (data from \cite{RamYD}).

\item[{\bf Fig. 2}] $\Delta I$ =2 staggering for the
rotational band  (2-2), $v=$1
of the
A$^1\Sigma^+$--X$^1\Sigma^+$ system of YN.
Even values of $I$ are used (data from \cite{RamYN}).

\item[{\bf Fig. 3}] $\Delta I$ =2 staggering for the
rotational band (4-3), $v$=1  for the  molecule CS.
Even values of $I$ are used (data from \cite{RamCS}).

\item[{\bf Fig. 4}] $\Delta I$ =4 staggering for the
superdeformed band (a) of the $^{149}$Gd nucleus.
(Data from \cite{Fli}.)

\item[{\bf Fig. 5}] $\Delta I$ =4 staggering for the
rotational band (1-1) , $v$=1 of the
  C$^1\Sigma^+$--X$^1\Sigma^+$ system of the molecule YD.
Even values of $I$ are used (data from \cite{RamYD}).

\end{itemize}

\end{document}